\documentstyle[aps,prl,twocolumn]{revtex}
\input amssym.def
\input amssym

\begin{document}
\title{Interferometry with Entangled Atoms}
\author{Ulvi Yurtsever}
\address{Quantum Computing Technologies Group, Jet Propulsion Laboratory,
California Institute of Technology
\newline
4800 Oak Grove Drive, Pasadena, California 91109-8099.}
\date{Received: \today}
\maketitle
\begin{abstract}
%\begin{center}
%\bf Abstract
%\end{center}
A quantum gravity-gradiometer
consists of two spatially separated ensembles of atoms interrogated
by pulses of a common laser beam. Laser pulses cause the
probability amplitudes of atomic ground-state hyperfine levels to
interfere, producing two motion-sensitive phase shifts which allow the
measurement of the average acceleration of each ensemble, and, via
simple differencing, of the acceleration gradient. Here I propose
entangling the quantum states of the two ensembles prior to the pulse sequence,
and show that entanglement encodes their relative acceleration in a single
interference phase which can be measured directly, with no need for differencing.

~~~~~

\end{abstract}
\parskip 3pt

~~~~~

{\noindent \bf 1. Introduction}

Inertial sensors based on matter-wave interferometry are
a rapidly developing technology [1---4]. One class of such sensors are
gradiometers designed to measure linear acceleration gradients, typically
gradients caused by inhomogeneities in the gravitational potential. The
state-of-the-art design\cite{Kasetal}
for an atom-wave gravity gradiometer consists of
a pair of atom-wave accelerometers separated spatially by a fixed distance and
direction. Each of the two accelerometers contains an ensemble of
laser-cooled atoms, through which a common pair of (counter-propagating)
Raman laser beams are pulsed in
a carefully controlled sequence to drive Rabi oscillations between
atomic ground-state hyperfine levels. The sequence
and timing of the laser pulses are adjusted so that the
hyperfine ground and excited states of any single atom interfere during
the atom's motion through the laser field. At the
end of the sequence, the probability of finding an atom in its excited
hyperfine state is a simple function of a relative phase, which
accumulates between successive pulses in
proportion to the atom's acceleration (relative to the inertial frame
defined by the laser beam) averaged over its flight time.
Subsequently, the resulting pair of observations of
this phase (one for each ensemble) can
be subtracted to obtain a measurement of the relative acceleration,
and, upon dividing the result by the separation length,
a measurement of the acceleration gradient.
I will give a more quantitative review of this quantum interferometry
technique in Sect.\,2 below.
A detailed description of the experimental setup
can be found in \cite{Kasetal} (see esp.\ Fig.\,1 there).

Even though subtracting the two acceleration measurements
allows a number of correlated noise sources to cancel out as common
modes\cite{Kasetal}, it is still desirable to avoid such differencing of two
nearly-equal measured quantities which are corrupted by (uncorrelated) noise.
In this letter, I will show that by prior-entangling (prior to the laser
pulsing sequence) the quantum states of atoms in
the two ensembles (such that every entangled pair has one atom in each ensemble)
it becomes possible to encode the relative-acceleration information directly in a
single interference phase, thereby eliminating the need for the
differencing of two separate phase measurements. This idea is inspired by
the recent discovery of a similar quantum algorithm which uses entangled
states to synchronize atomic clocks non-locally\cite{QCS}. The Quantum
Clock Synchronization (QCS) algorithm relies on preparing the atoms ``stabilizing"
each atomic clock in a special entangled quantum state whose time evolution
reduces to a pure multiplicative phase as long as each atom in the pair
evolves under the same unitary transformation. By contrast,
the Entangled Quantum Interferometry (EQI)
algorithm to be described below
relies on transforming the pair's quantum state into a sensor which
senses the {\em difference} between unitary evolutions of the entangled atoms.
Put another way, the EQI algorithm is a spatial analogue of
the QCS algorithm, in exactly the same sense as atomic clocks are
temporal analogues of quantum (atom-wave) interferometers\cite{ClockasInterf}.

{\noindent \bf 2. Overview of matter-wave gravity gradiometry}

Consider an atom in its internal
ground state $|0\rangle$, moving through the inertial frame defined by
the Raman beams of an atom interferometer.
[Here ``inertial frame" has the same meaning as in general relativity:
in the presence of a gravitational field, my coordinate system
will always be locally-Minkowski (free-falling, or geodesic-normal
coordinates) so that the gravitational potential and its
first derivatives vanish at the origin.]
The wave function can be written as a tensor product
of the atom's internal and external Hilbert-space states:
\begin{eqnarray}
\Psi_{0:\vec{p}} \, (\vec{x} , t) & = & | 0 \rangle \otimes \int 
a(\vec{k} , \omega ) \, e^{i (\vec{k} \cdot \vec{x} - \omega t)} \; d^3 k
\nonumber \\
& \equiv & | 0 \rangle \otimes \psi(\vec{x} , t ) \; ,
\end{eqnarray}
where $a(\vec{k} , \omega )$ defines the wave packet representing the external
(space-time) part $\psi(\vec{x} ,t )$
of the quantum state, $\omega$ is related to $\vec{k}$
via the usual dispersion relation (plus an additive constant corresponding
to the atom's internal energy level), and $\vec{p}\,$ denotes the
average (in general time-dependent) momentum:
\begin{equation}
\vec{p} \equiv \langle \psi | -i \hbar \nabla | \psi \rangle
= (2 \pi )^3 \int \hbar \vec{k} \;
| a(\vec{k} , \omega ) |^2 \; d^3 k \; .
\end{equation}
The need for this cumbersome subscript
notation $\Psi_{0:\vec{p}}$ will become clear in a moment.
Since the two Raman beams are detuned in frequency by an amount
$\Omega$ which corresponds to the energy difference between the ground
and excited states $|0\rangle$ and $|1\rangle$,
they might stimulate the ground-state atom to absorb a photon
from one beam and emit a lower frequency photon into the other, resulting
in a transition to the excited state $|1\rangle$.
This stimulated transition gives the atom a net momentum kick in
the amount $\hbar (\vec{k}_1 - \vec{k}_2) \equiv \hbar \vec{K}$,
where $\vec{k}_1$, $\vec{k}_2$ ($|\vec{k}_1| > |\vec{k}_2|$)
denote the wave vectors
corresponding to the (counter-propagating) Raman beams.
[As the vectors $\vec{k}_i$, $i=1,\; 2$, point in opposite directions,
the magnitude of $\vec{K}$ is typically twice that of the $\vec{k}_i$.
Note that $\Omega$ is the {\em total} energy absorbed by the atom,
which includes not only the transition energy (within the appropriate
bandwidth), but also the recoil contribution.]
The wave function of the same atom excited in this way from its ground state
can then be decomposed, similarly to Eq.\,(1), in the form
\begin{eqnarray}
\Psi_{1:\vec{p} + \hbar \vec{K}} \, (\vec{x} , t) & = & | 1 \rangle \otimes \int 
a(\vec{k} , \omega ) \, e^{i [(\vec{k} + \vec{K})
\cdot \vec{x} - (\omega + \Omega) t]} \; d^3 k
\nonumber \\
& = & | 1 \rangle \otimes \psi(\vec{x} , t ) \, e^{i (\vec{K}\cdot \vec{x}
- \Omega t )} \; .
\end{eqnarray}
Since the wave packet $\psi(\vec{x}, t)$ is common to both the
ground and excited state wave functions, I will suppress it (as an overall
``normalization" factor) in what follows.
Note that whenever an atom interrogated in the interferometer
is excited from its ground state $|0:\vec{p} \rangle$ to
the internal excited state $|1 \rangle$ by Raman pulses, it
always picks up an associated excess energy-momentum
$(\vec{K} , \Omega )$ as specified in Eq.\,(3); in other words,
the excited atom's total quantum state becomes
$|1:\vec{p}+\hbar \vec{K}\rangle$. Conversely, a stimulated transition
in the reverse direction results in the transformation
$|1:\vec{p}\rangle \longrightarrow |0:\vec{p}-\hbar\vec{K}\rangle$.
This correlation between an atom's internal state and its momentum
during the laser interrogation process is
the key feature enabling atom interferometry\cite{Chuetal}.

In the interferometer, atoms are manipulated by laser pulses
of two basic kinds: A Hadamard pulse $H_{\! \frac{\pi}{2}}$
(a $\pi /2$ pulse followed by the
spin operator $\sigma_z$), whose action is
\begin{eqnarray}
H_{\! \frac{\pi}{2}} |0:\vec{p}\rangle & = & \frac{1}{\sqrt{2}}
( |0:\vec{p}\rangle +
|1:\vec{p}+\hbar\vec{K}\rangle ) \; , \nonumber \\
H_{\! \frac{\pi}{2}} |1:\vec{p}+\hbar\vec{K}\rangle
& = & \frac{1}{\sqrt{2}}
( |0:\vec{p} \rangle - |1:\vec{p}+\hbar\vec{K} \rangle ) \; ,
\end{eqnarray}
and a ``double" Hadamard pulse $H_\pi$ (a $\pi$ pulse followed by the
spin operator $\sigma_z$), whose action is
\begin{eqnarray}
H_{\pi} |0:\vec{p} \rangle & = &
|1:\vec{p}+\hbar\vec{K} \rangle \; , \nonumber \\
H_{\pi} |1:\vec{p}+\hbar\vec{K} \rangle & = & |0:\vec{p} \rangle \; .
\end{eqnarray}
Consider a single atom's flight across the laser
field during the interrogation phase: Initially, at times $t<t_1$, say,
the atom is in the ground state $|0: \vec{p} \rangle$. At time $t=t_1$
the first $H_{\! \frac{\pi}{2}}$ pulse hits, and the atom's state
becomes (up to an overall phase factor which I will always ignore):
\begin{equation}
\Psi(t_1 ) = \frac{1}{\sqrt{2}} \, \left( \, |0:\vec{p}\rangle +
|1:\vec{p}+\hbar\vec{K}\rangle \, \right) \; .
\end{equation}
Since atoms are well-localized spatially, their
wave functions $\Psi(\vec{x}, t)$ are sharply peaked around their
average position $\langle \Psi (t)| \, \vec{x} \; | \Psi (t)\rangle$
at all times $t$. When an atom is excited by Raman pulses to a superposition
of its internal states, its
distribution becomes bimodal, but still sharply peaked around the two
positions (modes). Therefore, to a very good
approximation, I can follow the evolution of the atom's wave function
during its flight in the interferometer by examining it in the
vicinity of the atom's (average) position(s) as a function of
time $t$. (See \cite{Bimodality} for a more rigorous treatment.)
Accordingly, at any time $t$ after $t_1$ and before the next Raman
pulse hits, the state Eq.\,(6)
evolves, according to Eqs.\,(1)--(3), as
%\begin{eqnarray}
%\Psi(t) & = & \frac{1}{\sqrt{2}} \, \mbox{{\LARGE [}} \, |0:\vec{p}\rangle
%\nonumber \\
%& + &
%\left. e^{i\vec{K}\cdot (\vec{x}_t - \vec{x}_1 )
%- i \Omega (t - t_1)} |1:\vec{p}+\hbar\vec{K}\rangle \, \right] \; ,
%\nonumber \\
%& & \forall t: \;  t_1 \leq t < t_2  \; ,
%\end{eqnarray}
\begin{eqnarray}
\Psi(t) & = & \frac{1}{\sqrt{2}} \left[ |0:\vec{p}\rangle
+ e^{i\vec{K}\cdot (\vec{x}_t - \vec{x}_1 )
- i \Omega (t - t_1)} |1:\vec{p}+\hbar\vec{K}\rangle \, \right] ,
\nonumber \\
& & \forall t: \;  t_1 \leq t < t_2  \; ,
\end{eqnarray}
where $\vec{x}_t \equiv \vec{x} (t)$, $\vec{x}_1 \equiv \vec{x}(t_1)$
etc.\ is shorthand notation for the atom's position at the different
times, and an overall phase factor as well as the common wave
packet $\psi(\vec{x},t)$ are suppressed as advertised. At a time $t=t_2
> t_1$, an $H_\pi$ pulse hits, and the atom's new state becomes
\begin{eqnarray}
\Psi(t_2 )  & = & \frac{1}{\sqrt{2}} \,  \mbox{{\LARGE [}}
\, |1:\vec{p}+\hbar\vec{K}\rangle
\nonumber \\
& + &
\left. e^{i\vec{K}\cdot (\vec{x}_2 - \vec{x}_1 )
- i \Omega (t_2 - t_1)} |0:\vec{p}\rangle \, \right] \; .
\end{eqnarray}
At any time $t$ after $t_2$ and before the next pulse, the state
Eq.\,(8) evolves as
\begin{eqnarray}
\Psi(t) & = & \frac{1}{\sqrt{2}} \, \left[ \, 
e^{i\vec{K}\cdot (\vec{x}_t - \vec{x}_2 )
- i \Omega (t - t_2 )}
|1:\vec{p}+\hbar\vec{K}\rangle \right.
\nonumber \\
& + &
\left. e^{i\vec{K}\cdot (\vec{x}_2 - \vec{x}_1 )
- i \Omega (t_2 - t_1)} |0:\vec{p}\rangle \right] \; ,
\nonumber \\
& & \forall t: \;  t_2 \leq t < t_3  \; .
\end{eqnarray}
At time $t=t_3$, the second and last $H_{\! \frac{\pi}{2}}$ pulse hits,
and transforms $\Psi (t)$ into the state
\begin{eqnarray}
\Psi(t_3 ) & = & \frac{1}{2} \, \left[ \, 
e^{i\vec{K}\cdot (\vec{x}_3 - \vec{x}_2 )
- i \Omega (t_3 - t_2 )}
( |0:\vec{p}\rangle - |1:\vec{p}+\hbar\vec{K}\rangle ) \right.
\nonumber \\
& + &
\left. e^{i\vec{K}\cdot (\vec{x}_2 - \vec{x}_1 )
- i \Omega (t_2 - t_1)} ( |0:\vec{p}\rangle 
+ |1:\vec{p}+\hbar\vec{K}\rangle ) \right] .
\end{eqnarray}
So as a result of the pulse sequence ${H_{\! \frac{\pi}{2}}}
- {H_\pi}- H_{\! \frac{\pi}{2}}$, an atom which enters the
interferometer in the ground state $|0:\vec{p}\rangle$ ends up, at the
end of the last $H_{\! \frac{\pi}{2}}$ pulse (at $t=t_3$), in the state
Eq.\,(10) which can be rewritten as
\begin{eqnarray}
\Psi(t_3 ) & = & \frac{1}{2} \, \mbox{{\LARGE [}} \, 
\left( e^{i \Theta_{21}} + e^{i \Theta_{32}} \right)
|0:\vec{p}\rangle
\nonumber \\
& + &
\left. \left( e^{i \Theta_{21}} - e^{i \Theta_{32}} \right)
|1:\vec{p}+\hbar\vec{K}\rangle ) \right] ,
\end{eqnarray}
where for $i,j=1,2,3,\ldots$
\begin{equation}
\Theta_{ij} \equiv  \vec{K} \cdot (\vec{x}_i - \vec{x}_j ) - \Omega
(t_i - t_j ) \; .
\end{equation}
After absorbing a factor of $e^{i\Theta_{21}}$ into the overall phase
multiplying $\Psi(t_3)$, Eq.\,(11) takes the more familiar form
%\begin{equation}
%\Psi(t_3 ) = \frac{1}{2} \, \mbox{{\LARGE [}} \, 
%\left( 1 + e^{i \Theta} \right)
%|0:\vec{p}\rangle \nonumber
%+ \left( 1 - e^{i \Theta} \right)
%\left. |1:\vec{p}+\hbar\vec{K}\rangle ) \right] ,
%\end{equation}
\begin{eqnarray}
\Psi(t_3 ) & = & \frac{1}{2} \, \mbox{{\LARGE [}} \, 
\left( 1 + e^{i \Theta} \right)
|0:\vec{p}\rangle \nonumber \\
& + & \left( 1 - e^{i \Theta} \right)
\left. |1:\vec{p}+\hbar\vec{K}\rangle ) \right] ,
\end{eqnarray}
where $\Theta \equiv \Theta_{32} - \Theta_{21}$.
%\begin{equation}
%\Theta \equiv \Theta_{32} - \Theta_{21} \; .
%\end{equation}
The phase $\Theta$ can now be observed by measuring the relative
abundance of ground vs.\ excited state atoms in the state $\Psi(t_3 )$.
Specifically, the fraction $P_1$ of excited-state atoms in the state
Eq.\,(13) is given by
\begin{equation}
P_1 =\frac{1}{4} \left| 1 - e^{i \Theta} \right|^2 = \sin^2 
\left( \frac{\Theta}{2} \right) \; ,
\end{equation}
and, similarly, the fraction $P_0$ of ground state atoms is
$P_0 = | 1 + e^{i \Theta} |^2 /4 =  \cos^2 (\Theta /2)$. On the other
hand, substituting Eqs.\,(12) in
$\Theta = \Theta_{32} - \Theta_{21}$ gives
%\begin{eqnarray}
%\Theta & = & \vec{K} \cdot [ \vec{x}(t_3 ) - 2 \vec{x}(t_2 ) + \vec{x}(t_1 )]
%\nonumber \\
%& - & \Omega (t_3 - 2 t_2 + t_1 ) \; .
%\end{eqnarray}
\begin{equation}
\Theta = \vec{K} \cdot [ \vec{x}(t_3 )\! -\! 2 \vec{x}(t_2 )\! +\! \vec{x}(t_1 )]
- \Omega (t_3 \!- \! 2 t_2 \! +\! t_1 ) \; .
\end{equation}
If the Raman pulses making up the sequence ${H_{\! \frac{\pi}{2}}}
- {H_\pi}- H_{\! \frac{\pi}{2}}$ are aligned in time such
that $t_2 = t_1 + T$ and $t_3 = t_2 + T$ for some common interrogation
time $T$, the second term in parenthesis in Eq.\,(15) vanishes, and, using
the standard Taylor expansion of $\vec{x}(t)$, we can
rewrite the first term in the form
\begin{equation}
\Theta = \vec{K} \cdot \ddot{\vec{x}} (t_2 )
\, T^2 \; + \; O (T^3) \; .
\end{equation}
Here the magnitude of the $O(T^3 )$ remainder is down by a factor
of order $T/T_a$ relative to the first (acceleration) term,
where $T_a \sim \| \vec{a} \| / \| \dot{\vec{a}} \|$ denotes the timescale
over which the atom's acceleration
$\vec{a}(t)\equiv \ddot{\vec{x}}(t)$ varies.
Therefore, as long as the time scale $T_a$ is much larger (as is
typically the case) than the interrogation time $T$, measurement of the
phase $\Theta$ following the pulse sequence
$H_{\! \frac{\pi}{2}} \, {}_{\overline{\; T \; }} \,
{H_\pi} \, {}_{\overline{\; T \; }} \, H_{\! \frac{\pi}{2}}$
yields a direct measurement\cite{ftnote1} of the inertial acceleration
component $\vec{K} \cdot \vec{a}$.

{\noindent \bf 3. Interferometry with entangled atoms}

For a gradient measurement, ordinarily it is necessary to apply two separate but
simultaneous measurements of acceleration to two different ensembles
of atoms separated spatially by a fixed distance. By using atoms
whose internal states are pairwise entangled in a distributed
fashion, however, it is possible to obtain the acceleration gradient
information with just one phase measurement as I will show now:

Start, at time $t=t_1$,
with atoms in the two ensembles
$A$ and $B$ in the pairwise entangled initial state
\begin{eqnarray}
\Psi(t_1 ) & = & \frac{1}{\sqrt{2}} \, \left( \, |0:\vec{p}_A \rangle_A
\; |1:\vec{p}_B + \hbar\vec{K} \rangle_B \right. \nonumber \\
& - & \left.
|1:\vec{p}_A+\hbar\vec{K}\rangle_A \;
|0:\vec{p}_B \rangle_B \, \right) \; .
\end{eqnarray}
In other words, in some subset (which my measurements will be directed at) of
ensemble $A$, every atom is entangled with some atom in ensemble $B$, and,
at some initial time $t=t_1$, each entangled pair is (upto an overall phase)
in the state Eq.\,(17). Focus attention now on a single such entangled pair.
Since the state Eq.\,(17) is an energy eigenstate of the total internal
Hamiltonian, its explicit time evolution consists of a pure
multiplicative phase only. Implicitly, it evolves in time
kinematically with the inertial motion of the atoms; according to
Eqs.\,(1)--(3), this evolution takes the form
\begin{eqnarray}
\Psi(t) & = & \frac{1}{\sqrt{2}} \, \left[ \,
e^{i \vec{K} \cdot (\vec{x}_{Bt} - \vec{x}_{B1} )} \;
|0:\vec{p}_A \rangle_A
\; |1:\vec{p}_B + \hbar\vec{K} \rangle_B \right.
\nonumber \\
& - &
\left. 
e^{i \vec{K} \cdot (\vec{x}_{At} - \vec{x}_{A1} )} \;
|1:\vec{p}_A + \hbar\vec{K} \rangle_A
\; |0:\vec{p}_B \rangle_B
\right] \; ,
\nonumber \\
& & \forall t: \;  t_1 \leq t < t_2  \; ,
\end{eqnarray}
where $\vec{x}_{At} \equiv \vec{x}_A (t)$, $\vec{x}_{B1} \equiv
\vec{x}_B (t_1 )$, etc., and an overall phase factor as well as the common wave
packet $\psi_A (\vec{x}_A ,t) \psi_B (\vec{x}_B , t)$ are suppressed as
in Eq.\,(7) above. For increased readability, let me introduce the
following abbreviation which I will use throughout the rest of this letter:
for $E=A, \; B$,
%\begin{eqnarray}
%|0\rangle_A & \equiv & |0:\vec{p}_A \rangle_A \; ,
%\; \; \; \; \; \; 
%|1\rangle_A \equiv |1:\vec{p}_A+\hbar\vec{K} \rangle_A \; ,
%\nonumber \\
%|0\rangle_B & \equiv & |0:\vec{p}_B \rangle_B \; ,
%\; \; \; \; \; \; 
%|1\rangle_B \equiv |1:\vec{p}_B+\hbar\vec{K} \rangle_B \; .
%\end{eqnarray}
\begin{equation}
|0\rangle_E \equiv |0:\vec{p}_E \rangle_E \; ,
\; \; \; \; \; \; 
|1\rangle_E \equiv |1:\vec{p}_E+\hbar\vec{K} \rangle_E \; .
\end{equation}
Now suppose that, at a time $t= t_2 > t_1$,
a {\em common} $H_\pi$ pulse is applied to both atoms $A$ and $B$
(simultaneously) in the state Eq.\,(18).
Then the new state of the pair at $t=t_2$ becomes
\begin{eqnarray}
\Psi(t_2 ) & = & \frac{1}{\sqrt{2}} \, \left[ \,
e^{i \vec{K} \cdot (\vec{x}_{B2} - \vec{x}_{B1} )} \;
|1 \rangle_A
\; |0 \rangle_B \right.
\nonumber \\
& - &
\left. 
e^{i \vec{K} \cdot (\vec{x}_{A2} - \vec{x}_{A1} )} \;
|0 \rangle_A
\; |1 \rangle_B
\right] \; .
\end{eqnarray}
At any time $t$ after $t_2$ and before the next pulse, the state
Eq.\,(20) evolves as
\begin{eqnarray}
\Psi(t) & = & \frac{1}{\sqrt{2}} \, \left[ \,
e^{i \vec{K} \cdot (\vec{x}_{At} - \vec{x}_{A2}
+ \vec{x}_{B2} - \vec{x}_{B1} )} \;
|1 \rangle_A
\; |0 \rangle_B \right.
\nonumber \\
& - &
\left. 
e^{i \vec{K} \cdot (\vec{x}_{Bt} - \vec{x}_{B2}
+ \vec{x}_{A2} - \vec{x}_{A1} )} \;
|0 \rangle_A
\; |1 \rangle_B
\right] \; ,
\nonumber \\
& & \forall t: \;  t_2 \leq t < t_3  \; .
\end{eqnarray}
At a time $t=t_3 > t_2$, a final, {\em common} $H_{\! \frac{\pi}{2}}$
pulse is applied to the entangled pair $A$ and $B$, and transforms the
state Eq.\,(21) into
\begin{eqnarray}
\Psi(t_3 ) & = & \frac{1}{2 \sqrt{2}} \left[ 
e^{i ({\Phi^A}_{32} + {\Phi^B}_{21} ) }
(|0\rangle_A - |1 \rangle_A)
(|0 \rangle_B + |1\rangle_B ) \right.
\nonumber \\
& - &
\left. 
e^{i ({\Phi^B}_{32} + {\Phi^A}_{21} )}
(|0 \rangle_A + |1\rangle_A )
(|0\rangle_B - |1 \rangle_B )
\right] \; ,
\end{eqnarray}
where, for $i,j=1,2,3,\ldots$,
%\begin{eqnarray}
%{\Phi^A}_{ij} & \equiv & \vec{K} \cdot (\vec{x}_{Ai} - \vec{x}_{Aj} )
%\nonumber \\
%{\Phi^B}_{ij} & \equiv & \vec{K} \cdot (\vec{x}_{Bi} - \vec{x}_{Bj} ) \; .
%\end{eqnarray}
\begin{equation}
{\Phi^A}_{ij} \! \equiv \! \vec{K} \cdot (\vec{x}_{Ai} - \vec{x}_{Aj} ) ,
\; \;
{\Phi^B}_{ij} \! \equiv \! \vec{K} \cdot (\vec{x}_{Bi} - \vec{x}_{Bj} ) \; .
\end{equation}
After absorbing a factor of $e^{i ( {\Phi^B}_{32}+{\Phi^A}_{21})}$
into the overall phase
multiplying $\Psi(t_3)$, Eq.\,(22) takes the more manageable form
\begin{eqnarray}
\Psi(t_3 ) & = & \frac{1}{2 \sqrt{2}} \, \mbox{{\Large [}} \, 
e^{i \Phi}
(|0\rangle_A - |1 \rangle_A)
(|0 \rangle_B + |1\rangle_B ) 
\nonumber \\
& - &
(|0 \rangle_A + |1\rangle_A )
(|0\rangle_B - |1 \rangle_B )
\, \mbox{{\Large ]}} \; ,
\end{eqnarray}
with the phase observable $\Phi$ given by
\begin{eqnarray}
\Phi & \equiv & {\Phi^A}_{32}+{\Phi^B}_{21}-{\Phi^B}_{32}-{\Phi^A}_{21}
\nonumber \\
& = & \vec{K} \cdot (\Delta \vec{x}_3 - 2 \, \Delta \vec{x}_2 + \Delta
\vec{x}_1 ) \; ,
\end{eqnarray}
where $\Delta \vec{x} \equiv \vec{x}_A - \vec{x}_B$ and $\Delta \vec{x}_i
\equiv \vec{x}_{Ai} - \vec{x}_{Bi}$ for $i,j=1,2,3,\ldots \;$ Collecting
terms in Eq.\,(24),
\begin{eqnarray}
\Psi(t_3 ) & = & \frac{1}{2 \sqrt{2}} \, \mbox{{\Large [}} \, 
(e^{i \Phi} - 1)|0\rangle_A |0 \rangle_B
+ (e^{i \Phi} + 1) |0\rangle_A |1 \rangle_B 
\nonumber \\
& - &
(e^{i \Phi} + 1) |1 \rangle_A |0\rangle_B 
- (e^{i \Phi} - 1)|1\rangle_A |1 \rangle_B
\, \mbox{{\Large ]}} \; .
\end{eqnarray}
The phase $\Phi$ can now be observed by measuring the relative
abundance in the state $\Psi(t_3 )$ of those pairs in which both atoms
are in their excited states. [Even though $\Psi(t_3 )$ is not
stationary, its time evolution consists of pure phases multiplying each
of the coefficients in Eq.\,(26), so the observation time is not
critical.]
Specifically, according to Eq.\,(26), the fraction $P_{11}$ of
those pairs in which both the $A$-atom and the $B$-atom are in the excited
state $|1\rangle$ is given by
\begin{equation}
P_{11} =\frac{1}{8} \left| e^{i \Phi} -1 \right|^2 = \frac{1}{2} \sin^2 
\left( \frac{\Phi}{2} \right) \; .
\end{equation}
On the other hand,
if the Raman pulses making up the sequence $- {H_\pi}- H_{\! \frac{\pi}{2}}$
are aligned in time as before so that $t_2 = t_1 + T$ and $t_3 = t_2 + T$
for some common interrogation
time $T$, we can write, using Eq.\,(25),
\begin{equation}
\Phi = \vec{K} \cdot \Delta \ddot{\vec{x}} (t_2 )
\, T^2 \; + \; O (T^3) \; ,
\end{equation}
where the magnitude of the $O(T^3 )$ remainder is again down by a factor
of order $T/T_a$ relative to the first (acceleration) term,
with $T_a$ denoting the timescale over which the atoms' acceleration varies.
As long as the time scale $T_a$ is much larger than the interrogation time $T$,
measurement of the single phase observable
$\Phi$ following the pulse sequence
${}_{\overline{\; T \; }} \,
{H_\pi} \, {}_{\overline{\; T \; }} \, H_{\! \frac{\pi}{2}}$
on the entangled pair Eq.\,(17)
yields a direct measurement of the inertial acceleration gradient
$\vec{K} \cdot \Delta \vec{a} = \vec{K} \cdot (\vec{a}_A - \vec{a}_B )$
\nopagebreak{between the ensembles $A$ and $B$.
Signal sensitivity of $\Phi$ to the acceleration gradient $\Delta
\vec{a}$ can be enhanced using tricks similar to the one discussed in
the note\cite{ftnote1}.}

{\noindent \bf 4. Producing the entangled atom pairs}

A number of methods for entangling two-level atoms have been discussed
in the literature and some implemented in experiments. For
completeness, I will sketch two such methods which can be
used to produce states of the form Eq.\,(17).

One method, originally
proposed in Cabrillo {\it et.\ al.} \cite{Zoller},
involves pumping two spatially separated atoms $A$ and $B$
into their excited states, so that the joint system is initially in the
product state
\[
|1 \rangle_A |1 \rangle_B.
\]
A single-photon detector, which cannot
(even in principle) distinguish the atom
from which a detected photon arrives,
is placed halfway between the atoms. Such a detection scheme
can be implemented
either by using a detector which is inherently insensitive to the
direction of the photon that triggers it, or by using two detectors
situated at the output ports of a symmetric beam-splitter
on which the light paths from the two atoms are symmetrically
impingent, thus erasing the which-path information.
When one of the atoms makes a transition
to its ground state, and the detector registers the emitted photon,
the result of its measurement is to put the combined two-atom
system into the entangled state
\begin{equation}
\frac{1}{\sqrt{2}} \, \left( \, |0  \rangle_A
|1 \rangle_B + e^{i \phi}
|1 \rangle_A |0 \rangle_B \, \right) \; ,
\end{equation}
where the phase $\phi$ is a measure of the relative delay between the
paths reaching the detectors from $A$ versus from $B$, and can be adjusted
to have any desired value [e.g.\ as in Eq.\,(17)]
by adjusting the position of the detector
relative to the atoms.

A second method, recently investigated in detail by Haroche and
coworkers\cite{Haroche} is (in very rough outline)
the following: Start with a single-mode cavity whose excitation
frequency is tuned to $\Omega$. Send the pair of atoms $A$ and
$B$ into the cavity one after the other, with atom $B$ first.
Initially, both atoms and the cavity are in their ground states:
\begin{equation}
|0  \rangle_A \; |0  \rangle_B \; | 0 \rangle_{EM}
\;,
\end{equation}
where $| 0 \rangle_{EM}$ denotes the vacuum state of the cavity. After
atom $B$ is in the cavity, apply a $\pi/2$-pulse on it, which transforms
the state Eq.\,(30) into
\begin{equation}
\frac{1}{\sqrt{2}} |0  \rangle_A
\left( |0  \rangle_B \, | 1 \rangle_{EM}
- |1  \rangle_B |0 \rangle_{EM} \right) \; .
\end{equation}
When both atoms are in the cavity, apply a second,
$\pi$-pulse, this time on the atom $A$,
thereby transforming the state Eq.\,(31) into
\begin{equation}
\frac{1}{\sqrt{2}} \, \left( \,
|1  \rangle_A \, |0  \rangle_B 
\, - \, |0 \rangle_A \, |1 \rangle_B \, \right) \otimes |0 \rangle_{EM} \; ,
\end{equation}
which, for the atom-pair $A$ and $B$,
is in the desired form Eq.\,(17) up to an overall phase factor.
Considering that a number of proof-of-principle experiments
confirming it have already been
performed successfully\cite{Haroche},
this method (or, more precisely, a suitable variant which maintains
the necessary correlation between atomic energy levels and
linear momentum as discussed Sect.\,2 above) is likely to be the
method of choice in practical implementations of the entangled-atom
gradiometry algorithm.

{\noindent \bf 5. Quantifying the improvement}

To what extent does the entangled-interferometry technique
proposed in Sect.\,3 represent an improvement over the standard,
``classical" atom-wave gradiometry as reviewed in Sect.\,2?
To answer this question quantitatively, I will now carry out a
simple analysis of the improvement in the signal-to-noise
ratio expected with entangled-interferometry.

I will take the differential phase-shift as my observed signal,
so in a gradient measurement with both classical and entangled
interferometry, the signal is the dimensionless quantity
\begin{equation}
S = \vec{K} \cdot \Delta \vec{a} \; T^2 \; ,
\end{equation}
where $\vec{a} \equiv \ddot{\vec{x}}$, and other symbols are
defined as in Eqs.\,(16) and (28), with the same approximation that allows
neglecting terms of order $O(T^3 )$ being assumed implicitly.
In this discussion, for ease of notation I will label the two
atomic ensembles by $i=1, \;2$ as opposed to by $A$ and $B$
before (in Sects.\,2---3). In a
``classical" gradient measurement, two ensembles containing $M$ atoms
each are used to obtain the two acceleration measurements, which are later
differenced to obtain the signal
\begin{equation}
S = \Theta_1 - \Theta_2 \; ,
\end{equation}
where the phase observables $\Theta_i$ are related to the observed
abundance (fluorescence)
of excited-state atoms in each ensemble $i=1,\;2$ via
Eq.\,(14):
\begin{equation}
\sin^2 \left( \frac{\Theta_i}{2} \right) = \frac{M_i}{M} \; .
\end{equation}
Here $M_i$ denotes the number
of excited-state atoms at the end of the
interrogation process in ensemble $i$, $i=1,\; 2$.

For comparison, imagine a similar
experiment using the entangled algorithm of Sect.\,3, involving $M$
pairs of entangled atoms interrogated in the pulse sequence of
Eqs.\,(17)---(23) with the same interrogation-time constant
$T$. At the end of that sequence, the number of atom pairs
$M_{11}$ in which both atoms end up in their excited states
is measured, which yields the signal
\begin{equation}
S = \Phi \; ,
\end{equation}
where according to Eq.\,(27)
\begin{equation}
\frac{1}{2} \sin^2 \left( \frac{\Phi}{2} \right) = \frac{M_{11}}{M} \;.
\end{equation}

For simplicity, I will now make the additional assumption that the only
sources of noise in both kinds of signal measurements
(classical and entangled) are (i) phase noise, and (ii) shot noise
due to the statistical
fluctuations in the observed numbers of excited atoms:
in $M_1$ and $M_2$ for the classical case, and in $M_{11}$
for the entangled case. I will assume a binomial distribution for these
random variables with $M$ trials and with probabilities
$P_1$, $P_2$, and $P_{11}$, respectively. The expected RMS shot noise
in each number measurement $M_i$ is then given by
\begin{equation}
\Delta M_i \equiv 
\sqrt{<\! \! {M_i}^2 \! \! > - < \! \! M_i \! \! >^2}
= \sqrt{P_i (1 - P_i ) M } \; ,
\end{equation}
and the expected RMS noise in $M_{11}$ is, similarly,
\begin{equation}
\Delta M_{11} \equiv
\sqrt{< \! \! {M_{11}}^2 \! \! > - < \! \! M_{11} \! \! >^2}
= \sqrt{P_{11} (1 - P_{11} ) M } \; .
\end{equation}
In the classical measurement [Eqs.\,(34)---(35)], the shot
noise contribution to each $\Theta_i$
measurement can be written as
\begin{equation}
N_i = \frac{\partial \Theta_i}{\partial M_i} \Delta M_i \;
\; \; \; \; i=1,\; 2 \; .
\end{equation}
Since the two measurements of $\Theta_i$ are completely independent, the
total variance (noise squared)
in the quantity $S = \Delta \Theta$ is the sum of
the variances in each $\Theta_i$:
\begin{equation}
N_c = \left( {N_1}^2 + {N_2}^2 \right)^{\frac{1}{2}} \approx \sqrt{2}
N_1  = \sqrt{2} \, \frac{\partial \Theta_1}{\partial M_1} \Delta M_1\; ,
\end{equation}
where the approximate equality holds because of the
assumption that the signal $S \ll 1$. Substituting Eqs.\,(35), (38)
and (14) in Eq.\,(41) gives
\begin{equation}
N_c = \left( \frac{2}{M} \right)^{\frac{1}{2}} \; .
\end{equation}
In the entangled measurement, according to Eqs.\,(36)---(37)
\begin{equation}
S = \Phi = 2 \arcsin \sqrt{\frac{2 M_{11}}{M}} \approx
2\sqrt{2}\sqrt{\frac{M_{11}}{M}} \; ,
\end{equation}
and the shot noise contribution is
\begin{equation}
N_e = \frac{\partial \Phi}{\partial M_{11}} \Delta M_{11}
= \left( \frac{2 (1-
p_{11})}{M}\right)^{\frac{1}{2}} \approx
\left( \frac{2}{M} \right)^{\frac{1}{2}} \; ,
\end{equation}
where the second equality follows from Eq.\,(39), and the third
by the assumption that the signal $S \ll 1$. Remarkably, the shot
noise contributions in the classical and entangled measurements are
identical. Since a typical classical gradient measurement operates
in a regime where the
number of atoms $M$ in each ensemble $i$ is large enough to make shot
noise negligible\cite{Kasetal},
the same operating conditions will make it
negligible also in the entangled case, and therefore phase noise is the
dominant source of error in both kinds of measurement.

To compare phase-noise performance of the two kinds of measurement, I
will model the noise as simply proportional to the magnitude of
the fluorescence signal\cite{Kasetal}:
\begin{equation}
<\! \! {\Theta_i}^2 \! \! > - <\! \! \Theta_i \! \! >^2
= {\sigma}^2 \sin^2 \frac{<\! \! \Theta_i \! \! >}{2} \; , \;
\; \; \; i=1, \; 2 \; 
\end{equation}
for the classical case, and
\begin{equation}
<\! \! {\Phi}^2 \! \! > - <\! \! \Phi \! \! >^2
= {\sigma}^2 \frac{1}{2} \sin^2 \frac{<\! \! \Phi \! \! >}{2} \; , \;
\; \; \; i=1, \; 2 \; 
\end{equation}
in the entangled case, where $\sigma$ is a constant. By the same
arguments as above [Eqs.\,(40)---(41)], the phase noise
in the differential acceleration signal is, for the
classical case
\begin{equation}
N_c \approx \sqrt{2} \, \sigma \sin \frac{<\! \! \Theta_1 \! \! >}{2}
\; ,
\end{equation}
and for the entangled case
\begin{equation}
N_e = \frac{1}{\sqrt{2}} \, \sigma \sin \frac{<\! \! \Phi \! \! >}{2}
\; .
\end{equation}
Therefore the ratio of phase noise in the entangled measurement
to that in the classical measurement can be written in the form
[cf.\ Eqs.\,(36), (34), and (33)]
\begin{equation}
\frac{N_e}{N_c} = \frac{1}{2} \frac{\sin \frac{< \Phi >}{2}}
{\sin \frac{< \Theta_1 >}{2}} \approx
\frac{1}{2} \frac{<\! \! \Phi \! \! >}{<\! \! \Theta_1 \! \! >}
\sim \frac{1}{2} \frac{\| \Delta \vec{a} \|}{\| \vec{a} \|} \; .
\end{equation}
For a typical measurement performed on earth with a 1m separation
between the two atomic ensembles, the earth's $\sim \; 3000$E gradient
field (1E$ = 10^{-9} {\rm sec}^{-2}$) and 1g gravity suggests
suppression of phase noise via entanglement
by a factor of up to $1.5 \times 10^{-7}$, or more than six
orders of magnitude. For measurements performed in spacecraft in low
earth orbit, the ambient acceleration $\| \vec{a} \|$ will be
due to non-isolated common-mode vibrations originating from drag forces
acting on the spacecraft. Therefore, for a gradiometer
in low earth orbit (common-mode accelerations due
to atmospheric drag being
typically of order $10^{-5}$g or higher) the phase-noise suppression
$N_e / N_c$ via entangled interferometry can be expected to be a reduced
but still significant two orders of magnitude or better.

{\bf \noindent 6. Future work}

The Entangled
Quantum Interferometry algorithm as presented here can be
generalized to design sensors of other ``distributed" quantities (such
as higher derivatives of the gravitational curvature tensor); these
generalizations will be discussed in a future publication.

\begin{acknowledgements}
I would like to acknowledge valuable discussions with Lute
Maleki and Nan Yu of the JPL Frequency Standards Laboratory, and
with John Clauser, Jonathan Dowling, Dmitry Strekalov,
and Colin Williams of the JPL Quantum Computing Technologies group.
The research described in this paper
was carried out at the Jet Propulsion Laboratory,
California Institute of Technology, under a contract with
the National Aeronautics and
Space Administration, and was supported by a contract
with the National Reconnaissance Office.
\end{acknowledgements}


\begin{references}
\bibitem{ftnote1} Signal sensitivity in this
acceleration measurement can be improved by
employing more complicated pulsing sequences. For example, consider the
pulse sequence
\[
H_{\! \frac{\pi}{2}} \, {}_{\overline{\; \, 0  \; }} \,
{H_{(\pi )}} \, {}_{\overline{\; T \; }} \, {H_{(\pi )}}
\, {}_{\overline{\; \, 0  \; }} \,
{H_\pi} \, {}_{\overline{\; \, 0  \; }} \,
H_{(\pi )} \,  {}_{\overline{\; T \; }} \, 
H_{(\pi )} \, {}_{\overline{\; \, 0  \; }} \,
H_{\! \frac{\pi}{2}} \; ,
\]
where, for example, $H_{\! \frac{\pi}{2}} \, {}_{\overline{\; \, 0  \; }} \,
{H_{(\pi )}}$ denotes that the two successive
pulses are supposed to be applied one after the other
with zero time delay, and
$H_{(\pi )}$ denotes the ``minus-$\pi$" pulse defined by [cf.\ Eqs.\,(5)]
\begin{eqnarray}
H_{(\pi )} |0:\vec{p} \rangle & = &
|1:\vec{p}-\hbar\vec{K} \rangle \; , \nonumber \\
H_{(\pi )} |1:\vec{p} \rangle & = & |0:\vec{p}+\hbar\vec{K} \rangle
\; . \nonumber
\end{eqnarray}
(An $H_\pi$ pulse turns into to an $H_{(\pi )}$ pulse upon switching the
detuning order of the two Raman beams in such a way that the sign
flip $\vec{K} \longrightarrow - \vec{K}$ results.) Following through
a calculation similar to the one between Eqs.\,(6) and (13), it is
straightforward to show that an atom initially in the ground state
$|0: \vec{p}\rangle$ ends up, at the end of the longer pulse sequence
above, in a superposition state of the form Eq.\,(13)
where the phase $\Theta$ is now given by
$\Theta = 3 \, \vec{K} \cdot \vec{a}
\; T^2 + O (T^3)$. Hence a factor of 3 improvement in
the acceleration signal compared
to the sequence $H_{\! \frac{\pi}{2}} \, {}_{\overline{\; T \; }} \,
{H_\pi} \, {}_{\overline{\; T \; }} \, H_{\! \frac{\pi}{2}}$. [See,
for example, \cite{AI} (p.\ 388) and references cited therein
for a discussion on this and similar
techniques for boosting detection sensitivity.]
\end{references}
\end{document}